\documentclass[11pt,a4paper]{article}

\usepackage{epsfig}
\usepackage{amssymb}
\usepackage{graphicx}
\usepackage{color}
\usepackage{subfigure}

\usepackage{latexsym}

\begin{document}

\baselineskip 6mm
\renewcommand{\thefootnote}{\fnsymbol{footnote}}

\newcommand{\nc}{\newcommand}
\newcommand{\rnc}{\renewcommand}



\newcommand{\tcb}{\textcolor{blue}}
\newcommand{\tcr}{\textcolor{red}}
\newcommand{\tcg}{\textcolor{green}}


\def\beq{\begin{equation}}
\def\eeq{\end{equation}}
\def\ba{\begin{array}}
\def\ea{\end{array}}
\def\bea{\begin{eqnarray}}
\def\eea{\end{eqnarray}}
\def\nn{\nonumber}


\def\CMP{Commun. Math. Phys.~}
\def\JHEP{JHEP~}
\def\Pre{Preprint}
\def\PRL{Phys. Rev. Lett.~}
\def\PR {Phys. Rev.~}
\def\CQG {Class. Quant. Grav.~}
\def\PL {Phys. Lett.~}
\def\NP {Nucl. Phys.~}

\def\G{\Gamma}

\def\S{{\bf S}}
\def\C{{\bf C}}
\def\Z{{\bf Z}}
\def\R{{\bf R}}
\def\N{{\bf N}}
\def\M{{\bf M}}
\def\P{{\bf P}}
\def\bm{{\bf m}}
\def\bn{{\bf n}}

\def\CA{{\cal A}}
\def\CB{{\cal B}}
\def\CC{{\cal C}}
\def\CD{{\cal D}}
\def\CE{{\cal E}}
\def\CF{{\cal F}}
\def\CM{{\cal M}}
\def\CG{{\cal G}}
\def\CI{{\cal I}}
\def\CJ{{\cal J}}
\def\CL{{\cal L}}
\def\CK{{\cal K}}
\def\CN{{\cal N}}
\def\CO{{\cal O}}
\def\CP{{\cal P}}
\def\CQ{{\cal Q}}
\def\CR{{\cal R}}
\def\CS{{\cal S}}
\def\CT{{\cal T}}
\def\CV{{\cal V}}
\def\CW{{\cal W}}
\def\CX{{\cal X}}
\def\CY{{\cal Y}}
\def\We{{W_{\mbox{eff}}}}


\newcommand{\p}{\partial}
\newcommand{\bp}{\bar{\partial}}

\newcommand{\half}{\frac{1}{2}}

\newcommand{\bfalpha}{{\mbox{\boldmath $\alpha$}}}
\newcommand{\bfbeta}{{\mbox{\boldmath $\beta$}}}
\newcommand{\bfgamma}{{\mbox{\boldmath $\gamma$}}}
\newcommand{\bfmu}{{\mbox{\boldmath $\mu$}}}
\newcommand{\bfpi}{{\mbox{\boldmath $\pi$}}}
\newcommand{\bfvarpi}{{\mbox{\boldmath $\varpi$}}}
\newcommand{\bftau}{{\mbox{\boldmath $\tau$}}}
\newcommand{\bfeta}{{\mbox{\boldmath $\eta$}}}
\newcommand{\bfxi}{{\mbox{\boldmath $\xi$}}}
\newcommand{\bfkappa}{{\mbox{\boldmath $\kappa$}}}
\newcommand{\bfepsilon}{{\mbox{\boldmath $\epsilon$}}}
\newcommand{\bfTheta}{{\mbox{\boldmath $\Theta$}}}

\newcommand{\bz}{{\bar{z}}}

\newcommand{\dalpha}{\dot{\alpha}}
\newcommand{\dbeta}{\dot{\beta}}
\newcommand{\blambda}{\bar{\lambda}}
\newcommand{\btheta}{{\bar{\theta}}}
\newcommand{\bsigma}{{{\bar{\sigma}}}}
\newcommand{\bepsilon}{{\bar{\epsilon}}}
\newcommand{\bpsi}{{\bar{\psi}}}


\def\ct{\cite}
\def\la{\label}
\def\eq#1{(\ref{#1})}


\def\a{\alpha}
\def\b{\beta}
\def\g{\gamma}
\def\G{\Gamma}
\def\d{\delta}
\def\D{\Delta}
\def\ep{\epsilon}
\def\e{\eta}
\def\ph{\phi}
\def\Ph{\Phi}
\def\ps{\psi}
\def\Ps{\Psi}
\def\k{\kappa}
\def\l{\lambda}
\def\L{\Lambda}
\def\m{\mu}
\def\n{\nu}
\def\th{\theta}
\def\Th{\Theta}
\def\r{\rho}
\def\s{\sigma}
\def\S{\Sigma}
\def\ta{\tau}
\def\o{\omega}
\def\O{\Omega}
\def\pr{\prime}


\def\half{\frac{1}{2}}

\def\goto{\rightarrow}

\def\na{\nabla}
\def\grad{\nabla}
\def\curl{\nabla\times}
\def\div{\nabla\cdot}
\def\pa{\partial}

\def\bra{\left\langle}
\def\ket{\right\rangle}
\def\lb{\left[}
\def\lc{\left\{}
\def\ls{\left(}
\def\lp{\left.}
\def\rp{\right.}
\def\rb{\right]}
\def\rc{\right\}}
\def\rs{\right)}
\def\cl{\mathcal{l}}

\def\vac#1{\mid #1 \rangle}

\def\td#1{\tilde{#1}}
\def\check{ \maltese {\bf Check!}}


\def\Tr{{\rm Tr}\,}
\def\det{{\rm det}\,}


\def\bc#1{\nnindent {\bf $\bullet$ #1} \\ }
\def\ch {$<Check!>$ }
\def\ss {\vspace{1.5cm}}

\begin{titlepage}

\hfill\parbox{5cm} { }

\hskip1cm

\vspace{10mm}

\begin{center}
{\Large \bf Quasilocal Conserved Charges with a Gravitational Chern-Simons Term}

\vskip 1. cm
  {
  Wontae Kim$^{ab}$\footnote{e-mail : wtkim@sogang.ac.kr},  
  Shailesh Kulkarni$^{ac}$\footnote{e-mail : skulkarnig@gmail.com} and 
  Sang-Heon Yi$^{a}$\footnote{e-mail : shyi@sogang.ac.kr} 
  }

\vskip 0.5cm

{\it $^a\,$Center for Quantum Spacetime (CQUeST), Sogang University, Seoul 121-742, Korea}\\
{\it $^b\,$Department of Physics, Sogang University, Seoul 121-742, Korea}\\
{\it $^c\,$ Department of Physics, University of Pune, Ganeshkhind, Pune 411007, India}
\end{center}

\thispagestyle{empty}

\vskip1.5cm


\centerline{\bf ABSTRACT} \vskip 4mm

\vspace{1cm}
\noindent 
We extend our recent work on the quasilocal formulation of conserved charges to a theory of gravity containing a gravitational Chern-Simons term. As an application of our formulation, we compute the off-shell potential and   quasilocal conserved charges of some black holes in three-dimensional topologically massive gravity. Our formulation for conserved charges reproduces very effectively the well-known expressions on conserved charges and the entropy expression of black holes in the topologically massive gravity. 
\vspace{2cm}


\end{titlepage}

\renewcommand{\thefootnote}{\arabic{footnote}}
\setcounter{footnote}{0}

\section{Introduction}

Conserved charges in general relativity  is very important and rather subtle concept. The main obstacle is related to the construction of the completely generally covariant energy-momentum tensor for gravitational field. Many people, including Einstein himself,  have unsuccessfully tried to find such a tensor or  to construct its alternatives, for instance, energy momentum pseudo-tensor~\cite{landau: 1975}. Now there is a general consensus that such a construction does not exist,  since the local conservation law for general relativity turns out to be meaningless. However, several 
approaches have been suggested to construct total conserved charges in general relativity,  one of which is the formalism accomplished by Arnowit-Deser-Misner(ADM)~\cite{Arnowitt:1962hi}. 
This  approach uses a linearization of metric around the asymptotically flat spacetime and  becomes
cumbersome for the gravity actions  which contain higher curvature or higher derivative 
terms. An  extension of ADM formalism to higher curvature theories of gravity - known as the
Abbott-Deser-Tekin(ADT) formalism - was provided in \cite{Abbott:1981ff, Deser:2002jk}. 
Unlike the ADM formalism, the ADT method is covariant and also applicable to the asymptotically AdS geometry.

There also exist other approaches to conserved charges, which are based on quasilocal concepts  (for review, see \cite{Szabados:2004vb}). One of such formulations is the Brown-York formalism~\cite{Brown:1992br} which needs to be improved for asymptotic AdS space by introducing the appropriate counter terms~\cite{Balasubramanian:1999re}. This formulation has been especially useful in the context of the AdS/CFT correspondence.  Another such a formulation is known as the Komar integrals~\cite{Komar:1958wp}  which is not known to be completely consistent with the  results in the existing literatures. For instance, the mass and angular momentum calculated via Komar integrals contain the well-known factor two discrepancy when compared to ADM formalism.  In the covariant phase space approach, initiated by Wald  the conserved charges were computed by using the Noether potential~\cite{Wald:1993nt,Jacobson:1993vj, Iyer:1994ys, Wald:1999wa}. Wald's formulation has a distinct advantage in that it holds for any generally covariant theory of gravity and captures the entropy of black holes which can be regarded as  the natural extension of Beckestein-Hawking area law~\cite{Bekenstein:1973ur}. Furthermore, this method established  the first law of black hole thermodynamics in any covariant theory of gravity. 

There exists an interesting connection between the {\it on-shell} ADT potential and the linearized {\it on-shell} Noether potential.  Indeed, it was observed that at the asymptotic boundary, the linearized Noether potential around the on-shell background (which solves the Einstein equations), when combined with the surface term,
produces the known expression for the ADT potential \cite{Barnich:2001jy, Barnich:2003xg, Barnich:2004uw, Barnich:2007bf}.   
This relation, although very interesting, is indirect and shown to hold  in Einstein gravity only. 
In our recent work \cite{Kim:2013zha}, we have provided a non-trivial generalization
of the above connection to any covariant theory of gravity.  This was achieved by elevating
the ADT potential to the {\it off-shell}  level. Then, by using the corresponding {\it off-shell} 
expression for the linearized Noether potential supplemented with the surface term, we were able to show the desired connection directly. 
Integrating the resultant expression for the ADT potential along the one parameter path in the solution space,  we finally obtained the expression for the quasilocal conserved charges which is identical with the one given by Wald's covariant phase space approach.  At the on-shell level, this result establishes that our construction through  the quasilocal extension of the ADT formalism is completely equivalent to the covariant phase space formalism which encompasses  the black hole entropy.

There are certain aspects of our formalism which we would like to highlight at this stage. We may recall that in the conventional  analysis of ADT potentials/charges one has to use the  equations of motion for the background. As a result, the procedures become highly complicated when the higher curvature or higher derivative terms are present in the Lagrangian. On the other hand, our formalism uses the {\it off-shell} (or background independent)  expression for  ADT and Noether potentials which are shown to be related in a one to one fashion.  One can exploit this correspondence to obtain the conserved ADT charges for any covariant theory of gravity in a more efficient way. 
 The off-shell Noether potential has already been used  in the literatures in somewhat different ways. For instance, the entropy for black holes was computed from the {\it off-shell} Noether potential \cite{ Majhi:2011ws, Majhi:2012tf, Majhi:2012nq}. In another work \cite{Kim:2013qra}, we have used the {\it off-shell} Noether potential to compute the entropy for rotating extremal as well as non-extremal BTZ black holes in new massive gravity coupled to a scalar field. We have also computed the angular momentum of hairy AdS black holes and shown its invariance along the radial direction. This fact was used to verify the holographic c-theorem for hairy AdS black holes. 

Most of the studies on constructing the Noether potential and the corresponding conserved charges have been limited to  covariant theories of gravity. There are some attempts to generalize the Wald's formalism  to the apparently non-covariant Lagrangians which often include gravitational Chern-Simons terms. Gravitational Chern-Simons terms are closely related to anomaly and appear frequently in the string theory context. Moreover, it has important implications in the AdS/CFT correspondence.   
One of such a theory with a gravitational Chern-Simons term is the three-dimensional topologically massive gravity(TMG)  proposed by Deser, Jackiw and Templeton \cite{Deser:1981wh}. The extension of the Wald's  procedure to the TMG, especially for the black hole entropy, was provided by  Tachikawa in~\cite{Tachikawa:2006sz}. The entropy computed from this approach matches exactly with the one obtained by indirect ways  \cite{Cho:2004ey,Kraus:2005vz, Kraus:2005zm, Solodukhin:2005ah, Sahoo:2006vz, Park:2006gt}. This on-shell approach was extended in conjunction with the covariance of black hole entropy~\cite{Bonora:2011gz}.
On the other hand, the mass and angular momentum for the non-covariant theories like TMG
have been obtained independently of the entropy, for instance, by using the ADT formalism~\cite{Deser:2003vh, Bouchareb:2007yx}, by the canonical method~\cite{Hotta:2008yq} or by using the direct codified computer implementation~\cite{Chen:2013aza} of the formalism given  in~\cite{Barnich:2001jy}.  Another interesting aspect of TMG is the existence of warped AdS black hole solutions. These solutions with central charge expressions were a starting point for the warped AdS/CFT correspondence~\cite{Anninos:2008fx} and the Kerr/CFT correspondence~\cite{Guica:2008mu}, which may be extended to the dS/CFT case~\cite{Strominger:2001pn, Anninos:2009jt,deBuyl:2013ega}. 

Interestingly, the exact 
relationship among the ADT charges and the Noether potential is still missing for TMG.  
At first glance, the Noether potential  introduced in our previous paper \cite{Kim:2013zha} becomes non-covariant  for an apparently non-covariant Lagrangian like TMG. On the contrary, the ADT potential is completely covariant since its construction is based on the covariant equations of motion. Therefore, it is not clear that the formalism given in our previous paper can be extended to this case. In the present work we would expedite this apparently non-covariant case and show that the formalism works well even in this case. As a specific example, we will take the topologically massive gravity to elaborate our formalism. 

This paper is organized as follows.  In the next section we propose a general framework 
for calculating quasilocal conserved potentials and charges for a theory of gravity of the apparently non-covariant Lagrangian.   We then implement 
this procedure to TMG and show that our formalism matches completely with the covariant phase space approach to TMG. The quasi-local mass and angular momentum for the rotating Banados-Teitelboim-Zanelli(BTZ)~\cite{Banados:1992wn} and 
warped AdS black holes~\cite{Moussa:2003fc} are computed  in section 3. Finally, we summarize our findings in section 4.

\section{Conserved currents and potentials}
\subsection{Formalism}
In this section we extend our formulation of quasilocal conserved charges developed in~\cite{Kim:2013zha}. First, we obtain a generic expression for the off-shell Noether current.
This current, apart from the usual covariant terms, involves a non-covariant piece.  This means that the off-shell Noether current itself is not a covariant quantity and so does not have a good physical interpretation just like Levi-Civita connection. However, it turns out that its linearized expression is related to the off-shell covariant ADT potential and so it can be used for the computation of conserved charges through the one-parameter path on the  on-shell solution space. 

Let us consider an action which contains the apparently non-covariant term. The variation of the  action with respect to  $g^{\mu\nu}$ will be taken by
\beq 
\delta I = \frac{1}{16\pi G}\int d^{D}x~    \delta (\sqrt{-g} L ) =    \frac{1}{16\pi G}\int d^{D}x  \Big[  \sqrt{-g} \CE_{\mu\nu}\delta g^{\mu\nu}+\p_{\mu}\Theta^{\mu}(\delta g) \Big]\,, \label{genVar}
\eeq
where $\CE^{\mu\nu}=0$ denotes the  equations of motion (EOM) for the metric and $\Theta$ denotes the surface term. Note that the surface term $\Theta$ becomes non-covariant since we are considering the apparently non-covariant Lagrangian, though $\CE^{\mu\nu}$ is a covariant expression.
Under the diffeomorphism denoted by the parameter $\zeta$, the Lagrangian transforms as  
\beq
\delta_{\zeta} ( L \sqrt{-g}) =  \p_{\mu} (\sqrt{-g}\, \zeta^{\mu} L) +\p_{\mu}\Sigma^{\mu}(\zeta)\,,
\eeq
where $\Sigma^{\mu}$ term denotes an additional non-covariant term when the Lagrangian contains a non-covariant term like the  gravitational Chern-Simons term.

In this case, the identically conserved current can be introduced as
\beq J^{\mu} \equiv \p_{\nu}K^{\mu\nu}= \sqrt{-g}\, \CE^{\mu\nu}\zeta_{\nu} +  \sqrt{-g}\, \zeta^{\mu} L   + \Sigma^{\mu}(\zeta) - \Theta^{\mu}(\zeta) \,. \label{Noeth} \eeq
Unlike the covariant Lagrangian case, this off-shell Noether current $J^{\mu}$, and the potential $K^{\mu\nu}$ are not warranted to be covariant. This is naturally expected, since  the Lagrangian $L$, the surface term $\Theta$ and the boundary term $\Sigma$, all  take the non-covariant forms. Just as in the covariant case, there are some ambiguities in the form of the Noether potential $K$, which turn out not to affect the final expression for quasilocal conserved charges. 

In contrast, the ADT potential~\cite{Abbott:1981ff, Deser:2002jk, Deser:2002rt,Senturk:2012yi,Amsel:2012se}   is introduced in a completely covariant way. The on-shell ADT current is introduced for a Killing vector $\xi^{\mu}$ as  $\CJ^{\mu} = \delta \CE^{\mu\nu}\xi_{\nu}$, which can be shown to be conserved by using EOM, Bianchi identity and the Killing property of $\xi$. Then, the ADT potential $Q$ is introduced by $ \CJ^{\mu} = \nabla_{\nu}Q^{\mu\nu}$.
Since these on-shell current and potential, which use the EOM, are highly involved for a higher curvature or derivative theory of gravity, the background independent ADT current and potential were used for TMG~\cite{Bouchareb:2007yx} and new massive gravity~\cite{Nam:2010ub}. In Ref.~\cite{Kim:2013zha},  we have realized the importance of the identically conserved ADT current and extended its  use to a generic case.      Explicitly, the off-shell ADT current  and its potential  for a Killing $\xi$ can be defined by

\beq 
     {\cal J}^{\mu}_{ADT}  \equiv \nabla _{\nu}Q^{\mu\nu}_{ADT}= \delta \CE^{\mu\nu}\xi_{\nu} + \frac{1}{2}g^{\alpha\beta}\delta g_{\alpha\beta}\, \CE^{\mu\nu}\xi_{\nu}  + \CE^{\mu\nu}\delta g_{\nu\rho}\, \xi^{\rho}- \frac{1}{2}\xi^{\mu}\CE^{\alpha\beta}\delta g_{\alpha\beta} \,, \label{ADT}
\eeq
which can be shown to be conserved identically by using the Bianchi identity and the Killing property of $\xi$ without using EOM. Since this off-shell ADT potential is based on the covariant EOM, it takes the covariant form even for the apparently non-covariant Lagrangian.  
This covariant nature of the ADT potential  may lead some worries about the inapplicability of our formalism to the apparently non-covariant case. However as we shall see below,  the formalism can be extended successfully even to such a case.  

For matching the linearized off-shell Noether potential and the off-shell ADT potential,  the diffeomorphism parameter $\zeta$ will be taken as a Killing vector $\xi$ in the following. 
To extend the formalism in Ref.~\cite{Kim:2013zha} to this case,  let us introduce the formal Lie derivative for non-covariant $\Theta$ term as
\beq \CL_{\zeta} \Theta^{\mu} = \zeta^{\nu}\p_{\nu}\Theta^{\mu}  - \Theta^{\nu}\p_{\nu}\zeta^{\mu} + \p_{\nu}\zeta^{\nu}\, \Theta^{\mu}\,. \eeq
Note that this Lie derivative satisfies the Leibniz rule. This Lie derivative of a non-covariant quantity is different from its diffeomorphism variation, which is not the case for a covariant one. Let us denote  the difference between the Lie derivative and the diffeomorphism variation of  (non-covariant) $\Theta$-term as 
\beq  \delta_{\zeta} \Theta^{\mu}(g;\, \zeta) = \CL_{\zeta}\Theta^{\mu}(g;\, \delta g) + A^{\mu}(g;\, \delta g, \zeta)\,. \eeq
By using the property of the $\Theta$-term~\cite{Iyer:1994ys, Lee:1990nz}  for a Killing vector $\xi$
\beq \delta_\xi \Theta^{\mu}(g;\, \delta g) - \delta \Theta^{\mu}(g;\, \xi) =0\,, \eeq
and introducing $\Xi^{\mu\nu}$ as 
\beq    \p_{\nu} \Xi^{\mu\nu} (g;\, \delta g, \xi) \equiv  A^{\mu}(g;\, \delta g, \xi) -  \delta \Sigma^{\mu}(g;\, \xi)\,, \eeq
one can see that 
\beq 2\sqrt{-g}\, Q^{\mu\nu}_{ADT} = \delta K^{\mu\nu} (\xi)- 2 \xi^{[\mu}\Theta^{\nu]}(g;\, \delta g)  + \Xi^{\mu\nu}(g;\, \delta g, \xi)\,.  \label{Relation} \eeq
This is the extension of the quasilocal formula for conserved charges in the covariant Lagrangian case to the apparently non-covariant one.  The left hand side of the above equation is covariant by construction (see Eq.~(\ref{ADT})),  while each term in the right hand side is not warranted generically to be covariant. One may note that  the additional term $\Xi^{\mu\nu}$ is  responsible for the covariantization of the right hand side. We would like to emphasize again that this quasilocal ADT potential is defined only up to the total derivative of a certain antisymmetric tensor  $U^{\mu\nu\rho}$ just in the covariant case. This ambiguity does not affect the final expression for the conserved charges since it is a total derivative under the  integral over aclosed subspace. 

By using the above quasilocal ADT potential and using the one-parameter path in the solution space, just like the covariant case, one can introduce conserved charges for the Killing vector $\xi$ as
\beq  Q(\xi) = \frac{1}{8\pi G}\int^{1}_{0}ds \int d^{D-2} x_{\mu\nu} \sqrt{-g}\, Q^{\mu\nu}_{ADT}(g|s)\,. \eeq
We would like to emphasize that  the background and the variation are on-shell  in the end, since we have taken the path in the solution space.  The on-shell conservation of  $Q^{\mu\nu}_{ADT}$  has been used for construction of  conserved charges.  Using  Eq.~(\ref{Relation}), the conserved charge $Q(\xi)$ can be obtained through the Noether potential and surface terms as
\beq  Q(\xi) = \frac{1}{16\pi G} \int d^{D-2} x_{\mu\nu} \Big( \Delta K^{\mu\nu}(\xi) - 2\xi^{[\mu} \int^{1}_{0}ds\, \Theta^{\nu]} + \int^{1}_{0} ds\, \Xi^{\mu\nu}(\xi) \Big)\,,  \label{ConservedCharge}\eeq
where $\Delta K$ denotes the finite difference of $K$-values between two end points of the one-parameter path in the solution space.  The right hand side in Eq.~(\ref{ConservedCharge}) can be regarded as  the extension of the covariant phase space expression to the apparently non-covariant Lagrangian case, which was done at the on-shell level in~\cite{Tachikawa:2006sz}. On the bifurcate Killing horizon $H$, the second term in the right hand side would vanish and the final expression gives us the well-known Wald's entropy as $(\kappa/2\pi)\CS = Q_H$. Our construction shows that the conventional ADT charges should agree exactly with those from  the covariant phase  space approach. Conversely speaking, the quasilocal extension of the ADT formalism can reproduce the Wald's entropy for black holes. One of the lessons in this formulation is that the ADT charges and Wald's entropy do not need to be computed independently. Rather, they are directly related in our formulation and should be consistent with the first law of black hole thermodynamics by construction, as was shown by Wald~\cite{Wald:1993nt,Iyer:1994ys}.

\subsection{Gravitational Chern-Simons term}
In  this section we apply our formulation of quasilocal conserved charges to a specific example: TMG in three dimensions. It turns out that the ADT potential can be obtained in a very concise form and consistent with the previously known results. 
 
Let us take the action for TMG in three dimensions~\cite{Deser:1981wh} as 
\beq
 I [g_{\mu\nu}]  = \frac{1}{16\pi G}\int d^{3}x \sqrt{-g}\Big[R + \frac{2}{L^2} + \frac{1}{\mu}L_{CS} \Big]\,. \eeq
The last term for the gravitational Chern-Simons term is  given by
\beq
L_{CS} \equiv \frac{1}{2}\epsilon^{\mu\nu\rho}\Big(\G^{\alpha}_{\mu\beta}\p_{\nu}\G^{\beta}_{\rho\alpha} + \frac{2}{3}\G^{\alpha}_{\mu\beta}\G^{\beta}_{\nu\gamma}\G^{\gamma}_{\rho\alpha}\Big)\,, 
\eeq
where $\epsilon$-tensor is defined such that $\sqrt{-g}\, \epsilon^{012} =1$. Our convention for the curvature tensor is taken as $[\nabla_{\mu}, \nabla_{\nu}] V^{\rho}= R^{\rho}_{~\sigma\mu\nu}V^{\sigma}$ and the mostly plus metric signature is employed.  

The  equations of motion  for TMG are given by
\beq  G_{\mu\nu}  - \frac{1}{L^2} g_{\mu\nu}+  \frac{1}{\mu} C_{\mu\nu} =0\,, \eeq
where $G_{\mu\nu}$ denotes Einstein tensor and $C_{\mu\nu}$ denotes the Cotton tensor defined by
\beq  C^{\mu\nu} = \epsilon^{\alpha\beta\mu}\nabla_{\alpha}\Big(R^{\nu}_{\beta}  - \frac{1}{4}\delta^{\nu}_{\beta}R\Big)\,. \eeq
The above Cotton tensor is traceless, symmetric, and divergence-free, which is the three-dimensional analog of the Weyl tensor. One may note that it can also be written as $C^{\mu\nu} = \epsilon^{\alpha\beta(\mu}\nabla_{\alpha}R^{\nu)}_{\beta}$.  In the following, we use $h_{\mu\nu}$ for the linearized metric  interchangeably  with $\delta g_{\mu\nu}$ and all the indices are raised and lowered by the background metric $g$.

The quasilocal ADT potential for the Ricci scalar part has been known to be given by
\beq Q^{\mu\nu}_{R}(\xi) =   \frac{1}{2}h \nabla^{[\mu}\xi^{\nu]}  - h^{\alpha [\mu} \nabla_{\alpha} \xi^{\nu]}  -\xi^{[\mu} \nabla_{\alpha} h^{\nu]\alpha} + \xi_{\alpha}\nabla^{[\mu}h^{\nu]\alpha}  + \xi^{[\mu}\nabla^{\nu]} h\,, \eeq
which can also be derived from the quasilocal ADT formalism given in~\cite{Kim:2013zha}.
Since the construction has already been done for the covariant terms, 
let us focus on the gravitational Chern-Simons term in the following. The surface term for the gravitational Chern-Simons term  under a generic variation turns out to be
\beq \Theta^{\mu}(\delta g)= \frac{1}{2}\sqrt{-g}\Big[\epsilon^{\mu\nu\rho}\G^{\alpha}_{\rho\beta}\delta \G^{\beta}_{\nu\alpha} + \epsilon^{\alpha\nu\rho}R_{\nu\rho}^{~~\mu\beta}\delta g_{\alpha\beta}\Big]\,. \eeq
Note that the surface $\Theta$-term is non-covariant though the  EOM is covariant. 
Under a diffeomorphism with a parameter $\zeta$, Christoffel symbol transforms as
\beq \delta_{\zeta} \Gamma^{\rho}_{\mu\nu} = \CL_{\zeta} \Gamma^{\rho}_{\mu\nu}  + \p_{\mu}\p_{\nu}\zeta^{\rho}\,, \eeq
where $\CL_{\zeta}$ denotes the Lie derivative defined in the same way with the  $\Theta$-term as 
\[ \CL_{\zeta} \G^{\rho}_{\mu\nu} = \zeta^{\sigma}\p_{\sigma}\G^{\rho}_{\mu\nu} - \G^{\sigma}_{\mu\nu}\p_{\sigma}\zeta^{\rho} + 2 \G^{\rho}_{\sigma(\mu}\p_{\nu)}\zeta^{\sigma}\,.
\]
Then, one can see that $L_{CS}$ transforms under  diffeomorphism as
\beq \delta_{\zeta} L_{CS} = \p_{\mu}(\sqrt{-g}\zeta^{\mu} L_{CS} ) + \p_{\mu}\Sigma^{\mu}_{CS}\,, \eeq
where the additional boundary term $\Sigma_{CS}$ is given by
\beq \Sigma^{\mu}_{CS} = \frac{1}{2}\sqrt{-g}\, \epsilon^{\mu\nu\rho}\p_{\nu}\Gamma^{\beta}_{\rho\alpha}\, \p_{\beta}\zeta^{\alpha}\,. \eeq
The surface term for this diffeomorphism is given by
\beq  \Theta^{\mu}_{CS}(\zeta)= \sqrt{-g}\Big[\frac{1}{2}\epsilon^{\mu\nu\rho}\G^{\alpha}_{\rho\beta}\, \delta_{\zeta} \G^{\beta}_{\nu\alpha} +2 \epsilon^{\mu\nu(\alpha}R^{\beta)}_{\nu}\nabla_{\alpha}\zeta_{\beta}\Big]\,. \eeq
One may note that the above $\Sigma$-term and the $\Theta$-term have some ambiguities. Nevertheless, those do not affect our essential steps and so the above explicit expressions are taken for definiteness. 

According to the generic formulation given in Eq.~(\ref{Noeth}), the off-shell current and Noether potential  for a gravitational Chern-Simons term are introduced by
\beq J^{\mu}_{CS} \equiv \p_{\nu}K^{\mu\nu}_{CS} = 2\sqrt{-g}\,C^{\mu\nu} \zeta_{\nu} + \sqrt{-g}\,\zeta^{\mu}L_{CS} + \Sigma^{\mu}_{CS}(\zeta) - \Theta^{\mu}_{CS}(\zeta) \,. 
\eeq
By using three-dimensional identities given in the Appendix, one can obtain the off-shell Noether potential in the form of\footnote{See~\cite{Borowiec:1998st}  for  the on-shell Noether potential for a gravitational Chern-Simons term.}
\beq K^{\mu\nu}_{CS} = 2\sqrt{-g}\, \epsilon^{\mu\nu\rho} \bigg[ \Big(R^{\sigma}_{\rho} - \frac{1}{4}R\delta^{\sigma}_{\rho}\Big)\zeta_{\sigma}  -\frac{1}{4}\Gamma^{\beta}_{\rho\alpha} \nabla_{\beta}\zeta^{\alpha} \bigg]\,. \eeq
The additional term $\Xi^{\mu\nu}_{CS}$ can be shown to be given by
\bea  \Xi^{\mu\nu}_{CS}(g;\, \delta g, \zeta) &=& -\frac{1}{2}\sqrt{-g}\, \epsilon^{\mu\nu\rho}\delta \Gamma^{\beta}_{\rho\alpha}\p_{\beta}\zeta^{\alpha}  \\
&=&  -\frac{1}{2}\sqrt{-g}\, \epsilon^{\mu\nu\rho}\delta \Gamma^{\beta}_{\rho\alpha}\nabla_{\beta}\zeta^{\alpha}    + \frac{1}{2}\sqrt{-g}\, \epsilon^{\mu\nu\rho} \Gamma^{\alpha}_{\beta\sigma}\, \delta \Gamma^{\beta}_{\rho\alpha}\,\zeta^{\sigma}\,.   \nn 
\eea
Collecting the above results, one can obtain the contribution of the gravitational Chern-Simons term to conserved charges and the entropy of black holes. First, let us consider the contribution of the Chern-Simons term to the entropy of black holes. 
By using Eq.~(\ref{Relation}) with the on-shell background metric and taking $\zeta$  as the Killing vector $\xi_H$ for the Killing horizon $H$, one can show that the contribution of the  Chern-Simons term  is given by 
\beq  \frac{\kappa}{2\pi} \delta  \CS_{CS} = - \frac{1}{16\pi G} \int_{H} d^{D-2} x_{\mu\nu} \sqrt{-g}\, \epsilon^{\mu\nu\rho}\delta \Gamma^{\beta}_{\rho\alpha}\nabla_{\beta}\xi^{\alpha}_H  \,,   \eeq
where we have used the property of the bifurcate Killing horizon such that $\xi$ vanishes on $H$. This expression is completely covariant and  can be integrated into a finite form which is consistent with the one obtained in the covariant phase space approach~\cite{Tachikawa:2006sz,Bonora:2011gz}.

Now, by using our relation given in  Eq.(\ref{Relation}), one can obtain the quasilocal ADT potential for the three-dimensional gravitational Chern-Simons term as\footnote{We have been informed from B. Tekin that the essentially same expression was obtained in~\cite{Nazaroglu:2011zi}.}
\beq Q^{\mu\nu}_{CS} =  \epsilon^{\mu\nu\rho}\xi^{\sigma}\, \delta \Big(R_{\rho\sigma}- \frac{1}{4}Rg_{\rho\sigma} \Big)  - \frac{1}{2}\epsilon^{\mu\nu\rho}\delta \Gamma^{\beta}_{\rho\alpha}\nabla_{\beta}\xi^{\alpha}   - \xi^{[\mu}\epsilon^{\nu]\rho(\alpha} R^{\beta)}_{\rho}\delta g_{\alpha\beta} \,. 
\eeq
Note that this expression is completely covariant as was shown generically to be the case in the previous section. 
We would like to compare our results to the previously known expressions of the ADT potential for the gravitational Chern-Simons term. To achieve this goal, let us introduce the totally antisymmetric tensor $U^{\mu\nu\rho}$ as
\beq U^{\mu\nu\rho} \equiv  \frac{1}{2}\Big(\epsilon^{\rho\alpha\beta}\nabla_{\beta}\xi^{[\mu}h^{\nu]}_{\alpha} + \epsilon^{\alpha\beta[\mu}\nabla_{\beta}\xi^{\nu]}h^{\rho}_{\alpha} + h_{\alpha}^{[\mu}\epsilon^{\nu]\alpha\beta}\nabla_{\beta}\xi^{\rho} \Big)\,. \eeq
Using the Killing property of $\xi$ and the identities given in the Appendix, one can show that
$U^{\mu\nu\rho}$ satisfies 
\bea \nabla_{\rho}U^{\mu\nu\rho} &=& \frac{1}{2}hR\, \epsilon^{\mu\nu\rho}\xi_{\rho} + h\, R_{\alpha}^{[\mu}\epsilon^{\nu]\alpha\beta}\xi_{\beta}  +  R\, h_{\alpha}^{[\mu}\epsilon^{\nu]\alpha\beta}\xi_{\beta} -\frac{1}{2}\epsilon^{\mu\nu\rho}\xi_{\rho}\, h^{\alpha\beta}R_{\alpha\beta}   \nn \\
&& + \epsilon^{\alpha\beta\rho}\, h_{\alpha}^{[\mu}R^{\nu]}_{\beta}\xi_{\rho} - h_{\alpha\beta}R^{\alpha[\mu}\epsilon^{\nu]\beta\rho}\xi_{\rho} - R_{\alpha\beta}h^{\alpha[\mu}\epsilon^{\nu]\beta\rho}\xi_{\rho}\,. \nn
\eea
As a result, one can verify that the above expression of the ADT potential for the gravitational Chern-Simons term $Q^{\mu\nu}_{CS}$ can be rewritten as\footnote{see Appendix for some details of this computation.}
\beq Q^{\mu\nu}_{CS} = \nabla_{\rho}U^{\mu\nu\rho} + Q^{\mu\nu}_{R}(\eta) + \epsilon^{\mu\nu\rho}\Big[\delta G^{\lambda}_{\rho}\xi_{\lambda} - \frac{1}{2}\delta G \xi_{\rho}  + \frac{1}{2}\xi_{\rho}h^{\alpha\beta}G_{\alpha\beta} + \frac{1}{4}h(\xi_{\sigma}G^{\sigma}_{\rho} + \frac{1}{2}\xi_{\rho}R) \Big]\,,
\eeq
where $\eta$ is defined by $\eta^{\mu} \equiv  \frac{1}{2}\epsilon^{\mu\alpha\beta}\nabla_{\alpha}\xi_{\beta}.$
This computation shows us explicitly the equivalence of our expression of the background independent or {\it off-shell} ADT potential to the one given in~\cite{Bouchareb:2007yx}. Though our expression of the off-shell ADT potential is more succinct and illuminating,  we would like to emphasize that we can use  Eq.~(\ref{ConservedCharge}) for the computation of conserved charges instead of the explicit expression of the  ADT potential.  Using Eq.~(\ref{ConservedCharge}), one can  also obtain  the entropy of black holes in TMG at one stroke.

In order to apply Eq.~(\ref{ConservedCharge}) to  black hole solutions in TMG in the next section,  let us summarize what we have computed.   In TMG, the  off-shell Noether potential, $\Theta$-term,  and $\Xi$-term  are given by
\bea 
K^{\mu\nu}_{TMG}(g;\, \zeta) &=& \sqrt{-g}\bigg[ 2\nabla^{[\mu}\zeta^{\nu]} + \frac{2}{\mu}  \epsilon^{\mu\nu\rho} \Big\{ \Big(R^{\sigma}_{\rho} - \frac{1}{4}R\delta^{\sigma}_{\rho}\Big)\zeta_{\sigma}  -\frac{1}{4}\Gamma^{\beta}_{\rho\alpha} \nabla_{\beta}\zeta^{\alpha} \Big\} \bigg]\,,  \label{eq:potentials} \\
\Theta^{\mu}_{TMG}(g, \delta g) &=&  \sqrt{-g}\bigg[ \nabla^{\mu}(g_{\alpha\beta}\delta g^{\alpha\beta}) -\nabla_{\nu}\delta g^{\mu\nu}  + \frac{1}{\mu}  \Big\{\frac{1}{2}\epsilon^{\mu\nu\rho}\G^{\alpha}_{\rho\beta}\, \delta \G^{\beta}_{\nu\alpha} + \epsilon^{\mu\nu(\alpha}R^{\beta)}_{\nu}\delta g_{\alpha\beta}\Big\} \bigg]\,,  \nn \\
\Xi^{\mu\nu}_{TMG} (g;\, \delta g, \zeta)  &=& -\frac{1}{2\mu}\sqrt{-g}\, \epsilon^{\mu\nu\rho}\delta \Gamma^{\beta}_{\rho\alpha}\p_{\beta}\zeta^{\alpha}\,.    \nn  \eea
It is interesting to note that each of the above expressions are non-covariant, as expected. 

\section{Black holes and their charges}
In this section, we compute the mass and angular momentum of some black holes in TMG as the simplest example of our formulation. Since our formulation was shown to give us  the background independent ADT potential  which is equivalent to the previously known expression in~\cite{Bouchareb:2007yx}, the  mass and  angular momentum for black holes\footnote{Since the computation of the entropy of these black holes is identical with the covariant phase space approach and gives nothing new, we omit these parts.}  in TMG are assured to be given by the same expression. However, it is illuminating and fruitful to reproduce these results by using the expression of conserved charges given in~Eq.(\ref{ConservedCharge}).  
In all the given examples,  upper index components of relevant Killing vectors are taken to be constant and so $\Xi$-term contribution vanishes. 

In our convention the metric of the BTZ black hole~\cite{Banados:1992wn} is taken in  the following form of
\beq ds^2 = L^2\bigg[ -\frac{(r^2 - r^2_+)(r^2 - r^2_-)}{r^2}dt^2 + \frac{r^2}{(r^2 - r^2_+)(r^2 - r^2_-)}dr^2 + r^2\Big(d\theta - \frac{r_+r_-}{r^2}dt\Big)^2\bigg]\,. \eeq
The Killing vectors for the time-translational and rotational  symmetry will be chosen as $\xi =\frac{\p}{L\p t}, \frac{\p}{\p\theta}$, respectively.  To utilize the formula given in~Eq.(\ref{ConservedCharge}), take an infinitesimal parametrization of a one-parameter path in the solution space as follows
\[    r_+ \longrightarrow r_+ + dr_+\,, \qquad r_+ \longrightarrow r_- + dr_-\,. \]
By expanding the above BTZ  metric in terms of $dr_{\pm}$ and keeping terms up to the relevant order, one can obtain the  infinitesimal expression of  the $\Theta$-term. And then, one can integrate this expression to obtain conserved charges.  Let us consider the quasilocal angular momentum of the BTZ black hole at first. After a bit of computation, one can see that, just like the covariant case, the quasilocal  angular momentum  for the rotational Killing vector $\xi_R = \frac{\p}{\p \theta}$ comes entirely from the  $\Delta K$-term, of which  the relevant component  is 
\[
\Delta K^{r t}_{TMG}(\xi _R) = - 2 L r_+ r_-  + \frac{1}{\mu}(r^2_+  + r^2_-)\,.  
\]
As a result, the angular momentum of the BTZ black hole is given by
\beq J =  \frac{1}{16\pi G} \int^{2\pi}_{0}d\theta~  \Delta K^{r t}_{TMG} = - \frac{~ Lr_+r_-}{4 G} + \frac{r^2_+  + r^2_-}{8G\mu} \,. \eeq 
By noting that the nonvanishing components of the infinitesimal $\Theta$-term and the $\Delta K$-term for a Killing vector $\xi_T = \frac{1}{L}\frac{\p}{\p t}$ are   
\[ \Theta^{r} = L\,  d(r^2_+ +  r^2_-)\,, \qquad  \Delta K^{r t}(\xi_T) = -2\frac{r_+r_-}{L\mu}\,,
 \]
one can show that the mass of the BTZ black hole in TMG is given in the form of
\beq
 M = \frac{1}{16\pi G} \int^{2\pi}_{0}d\theta~ \Big( \Delta K^{r t}_{TMG}(\xi_T)  +  \xi^t_T  \int \, \Theta^{r} \Big)= \frac{\, r^2_+ + r^2_-}{8 G} - \frac{r_+r_-}{4GL\mu}\,.
\eeq
These expressions match completely with the known results. Note that our convention is such that the first law of black hole thermodynamics holds in the form of $dM = T_Hd\CS_{BH} - \Omega dJ$.

Now, let us consider the warped AdS black hole in TMG, of which expressions for the mass and angular momentum are rather involved.  The metric of the warped AdS black hole may be taken as~\cite{Moussa:2003fc}
\beq ds^2 = - \beta^2\frac{\rho^2 - \rho^2_0}{Z^2}dt^2 + \frac{d\rho^2}{\zeta^2\beta^2(\rho^2 -\rho^2_0)} + Z^2\Big(d\theta - \frac{\rho + (1-\beta^2)\omega}{Z^2}dt\Big)^2\,, \eeq
where $Z^2 \equiv \rho^2 + 2\omega \rho + (1-\beta^2)\omega^2 + \beta^2\rho^2_0/(1-\beta^2)$. Two of the four parameters in the above metric, $\beta$ and $\zeta$, are related to the Lagrangian parameter $1/L^2$ and $1/\mu$ as follows
\beq  \beta^2 \equiv  \frac{1}{4}\Big(1 + \frac{27}{\mu^2L^2}\Big)\,, \qquad \zeta = \frac{2}{3}\mu\,.  \eeq
The other two parameters $\omega$ and $\rho_0$ are related to the mass and angular momentum of this black hole.  In this case  one can choose the  infinitesimal one-parameter path  in the solution space as\footnote{Note that we do not need to expand in terms of $d\rho_0$ since the form of $\rho^2_0$ only appears in the metric.}  
\[ \omega \longrightarrow \omega + d \omega\,, \qquad \rho^2_0 \longrightarrow \rho^2_0 + d\rho^2_0\,. \]
As in the case of  the BTZ black hole, it is sufficient to   keep various terms up to linear parts of the variations.   Then, the quasilocal conserved angular momentum  for the rotational Killing vector $\xi_R = \frac{\p}{\p \theta}$  can be shown to come  entirely from the $\Delta K^{\mu\nu}$-term, while the quasilocal mass for the timelike Killing vector $\xi_T = \frac{\p}{\p t}$ has another contribution from the $\Theta$-term.  

Let us consider the angular momentum of the warped AdS black hole at first. By using the relevant component of the $\Delta K$-term for the rotational Killing vector $\xi_R$ 
\beq \Delta K^{\rho t}_{TMG}(\xi_R) = - \frac{2}{3}\zeta\beta^2 \Big[(1-\beta^2)\omega^2 - \frac{1+\beta^2}{1-\beta^2}\rho^2_0\Big]\,,  \eeq
one can obtain the quasilocal angular momentum of the warped $AdS_3$ black hole as
\beq J = \frac{1}{16\pi G} \int^{2\pi}_{0}d\theta~  \Delta K^{\rho t}_{TMG} (\xi_R)= -\frac{\zeta\beta^2}{12G}\Big[(1-\beta^2)\omega^2 - \frac{1+\beta^2}{1-\beta^2}\rho^2_0\Big]\,. \eeq
Now, let us turn to the mass of the black hole for the timelike Killing vector $\xi_T$.
In this case the nonvanishing component of the infinitesimal $\Theta$-term for the above chosen path turns out to be
\beq  \Theta^{\rho}_{TMG}   = \frac{2}{3}\zeta\beta^2(1-\beta^2)~ d\omega\,. \eeq
By combining this with  the $\Delta K$ contribution 
\beq
\Delta K^{\rho t}_{TMG}(\xi_T) = \frac{2}{3}\zeta \beta^2(1-\beta^2)\omega\,, \qquad \eeq
one can see that mass is given by
\beq M = \frac{1}{16\pi G} \int^{2\pi}_{0}d\theta~ \Big( \Delta K^{\rho t}_{TMG} (\xi_T) +  \xi^t_T  \int \, \Theta^{\rho} \Big)= \frac{\zeta\beta^2}{6G}(1-\beta^2)\omega\,. \eeq
Note that the above expressions for the mass and angular momentum match  completely with those given in~\cite{Bouchareb:2007yx} up to sign convention for angular momentum. (See~\cite{Anninos:2008fx,Compere:2008cv, Ghodsi:2011ua} for a dual CFT interpretation for these black holes.)

\section{Conclusion}
In this paper we have extended our previous formalism for quasilocal conserved charges to a theory of gravity with a gravitational Chern-Simons term. This formulation turns out to be very effective to obtain the ADT potential and quasilocal charges. In fact, we have shown that this quasilocal extension of the ADT method even to an apparently non-covariant Lagrangian is completely equivalent to the covariant phase space approach.  We have explicitly verified that this formulation reproduces the known background independent ADT potential for TMG up to the irrelevant total derivative of a totally antisymmetric tensor. Furthermore, quasilocal conserved charges for the  BTZ black holes and the warped AdS black holes are reproduced, which are completely consistent with the previously known results. 

It would be very interesting to develop this formulation further to encompass the asymptotic Killing vectors, which is relevant to the construction of the asymptotic Virasoro algebra in the context of the AdS/CFT, Kerr/CFT and dS/CFT correspondence. This would allow us to extract the information of the 
central charge and eventually the black hole entropy. Another interesting direction would be to use the {\it off-shell} Noether potential, $K_{\mu\nu}$ (see Eq.~(\ref{eq:potentials})) in the stretched horizon approach developed by Carlip~\cite{Carlip:1999cy}. This will lead to the near horizon Virasoro algebra and the entropy of black holes.\\

 \section*{Acknowledgments}
We would like to thank B. Tekin for  a useful correspondence. 
This work was supported by the National Research Foundation of Korea(NRF) grant funded by the Korea government(MSIP) through the CQUeST of Sogang University with grant number 2005-0049409. W. Kim was supported by the National Research Foundation of Korea(NRF) 
grant funded by the Korea government (MOE) (2010-0008359). S.-H.Yi was supported by the National Research Foundation of Korea(NRF) grant funded by the Korea government(MOE) (No.  2012R1A1A2004410). 
S. Kulkarni was also supported by the INSPIRE faculty scheme (IFA-13 PH-56)  by the Department of Science 
and Technology (DST), India.
\newpage
\quote{\bf\Large Appendix}\\ \\
\renewcommand{\theequation}{A.\arabic{equation}}
  \setcounter{equation}{0}
%
Here we shall give some identities and formulae which are useful in the text, especially in Section 2.2. In three dimensions we have the following identities 
\bea \epsilon^{\mu\nu\rho}V^{\sigma} &=& g^{\mu\sigma}\epsilon^{\nu\rho\alpha} V_{\alpha}+ g^{\nu\sigma}\epsilon^{\rho\mu\alpha}V_{\alpha} +g^{\rho\sigma}\epsilon^{\mu\nu\alpha}V_{\alpha}\,,   \\ 
R_{\mu\nu\rho\sigma} &=& R_{\mu\rho}g_{\nu\sigma}  + R_{\nu\sigma} g_{\mu\rho} - R_{\mu\sigma}g_{\nu\rho} - R_{\nu\rho}g_{\mu\sigma} - \frac{1}{2}R(g_{\mu\rho}g_{\nu\sigma} - g_{\nu\rho}g_{\mu\sigma})\,. \nn \eea
We have used the following convention for $\epsilon$ tensor and the integration measure 
\beq  \sqrt{-g}\, \epsilon^{tr\theta} = 1\,, \qquad  dx_{\mu\nu} \equiv  dx^{\rho}\, \epsilon_{\mu\nu\rho}\frac{1}{2\sqrt{-g}}\,. \eeq
A Killing vector $\xi$ satisfies 
\beq \nabla_{(\mu}\xi_{\nu)} =0\,, \qquad \nabla_{\mu}\nabla_{\nu}\xi_{\rho} = \xi^{\sigma}R_{\sigma\mu\nu\rho}\,. \eeq
For a Killing vector $\xi$, let us introduce another vector field $\eta$ formed by contracting the covariant derivative of $\xi$ with the $\epsilon$-tensor
\beq \eta^{\mu} \equiv \frac{1}{2}\epsilon^{\mu\alpha\beta}\nabla_{\alpha}\xi_{\beta}\,. \eeq
Such a vector field $\eta$ obeys  
\bea  \nabla^{[\mu}\eta^{\nu]} &=& \frac{1}{2}R\epsilon^{\mu\nu\rho}\xi_{\rho}+  R_{\alpha}^{[\mu}\epsilon^{\nu]\alpha\beta}\xi_{\beta} \,,   \\ && \nn \\\
-h^{\alpha[\mu}\nabla_{\alpha}\eta^{\nu]} &=& \frac{1}{2}Rh_{\alpha}^{[\mu}\epsilon^{\nu]\alpha\beta}\xi_{\beta} + \epsilon^{\alpha\beta\rho}h_{\alpha}^{[\mu}R_{\beta}^{\nu]}\xi_{\rho}\,, \nn \\ && \nn \\ 
\eta^{[\mu} \nabla_{\alpha}h^{\nu]\alpha} &=& \frac{1}{2} \nabla_{\rho}\xi_{\sigma}\, \epsilon^{\rho\sigma[\mu}\nabla_{\alpha}h^{\nu]\alpha}\,, \nn \\   && \nn \\ 
\eta^{[\mu}\nabla^{\nu]}h &=& \frac{1}{2} \nabla_{\rho}\xi_{\sigma}\, \epsilon^{\rho\sigma[\mu}\nabla^{\nu]}h\,,  \nn \\  && \nn \\
-\frac{1}{2}\epsilon^{\mu\nu\rho}\,\delta \Gamma^{\beta}_{\rho\alpha}\nabla_{\beta}\xi^{\alpha} &=& \eta_{\alpha}\nabla^{[\mu}h^{\nu]\alpha}  - \eta^{[\mu}\nabla_{\alpha}h^{\nu]\alpha}  + \eta^{[\mu}\nabla^{\nu]}h\,, \nn  \\ && \nn \\
 \eta_{\alpha}\nabla^{[\mu}h^{\nu]\alpha} &=& \frac{1}{2}\nabla_{\rho}\xi_{\sigma}\, \epsilon^{\rho\sigma\alpha}\nabla^{[\mu}h^{\nu]}_{\alpha}  = \epsilon^{\alpha\rho\beta}\nabla_{\rho}\xi^{[\mu}\nabla_{\beta}h^{\nu]}_{\alpha} \nn \\
&=&  2h \nabla^{[\mu}\eta^{\nu]}  + \frac{5}{2}Rh^{[\mu}_{\alpha}\epsilon^{\nu]\alpha\rho}\xi_{\rho} - \epsilon^{\mu\nu\rho}\xi_{\rho}\, h^{\alpha\beta}R_{\alpha\beta} \nn \\ 
&& - 2h^{\beta}_{\alpha}R^{[\mu}_{\beta}\epsilon^{\nu]\alpha\rho}\xi_{\rho} -  3 R^{\rho}_{\alpha}\, h^{[\mu}_{\rho}\epsilon^{\nu]\alpha\rho}\xi_{\rho} + 2\epsilon^{\alpha\beta\rho}h^{[\mu}_{\alpha}R^{\nu]}_{\beta}\xi_{\rho} \nn \\ 
&&  
 -\nabla_{\beta}\Big(\epsilon^{\alpha\rho[\mu}\nabla_{\rho}\xi^{\nu]}h^{\beta}_{\alpha} + h_{\alpha}^{[\mu}\epsilon^{\nu]\alpha\rho}\nabla_{\rho}\xi^{\beta} \Big)\,.  \nn  
\eea
Here,  $h_{\mu\nu}$ and $h$ represents the linearized metric and its trace, respectively. Another useful identity  for the $\nabla_{\rho} U^{\mu\nu\rho}$ computation is 
\bea
\nabla_{\beta}(\epsilon^{\alpha\rho\beta}\nabla_{\rho}\xi^{[\mu}h^{\nu]}_{\alpha}) &=& \epsilon^{\alpha\rho\beta}\nabla_{\rho}\xi^{[\mu}\nabla_{\beta}h^{\nu]}_{\alpha} + \frac{1}{2}Rh^{[\mu}_{\alpha}\epsilon^{\nu]\alpha\rho}\xi_{\rho} + R^{[\mu}_{\alpha}h^{\nu]}_{\beta}\, \epsilon^{\alpha\beta\rho}\xi_{\rho}  \nn \\
&& + \xi_{\sigma}R^{\sigma}_{\rho}\epsilon^{\alpha\rho[\mu}h^{\nu]}_{\alpha}\,.  \nn 
\eea

\newpage



\begin{thebibliography}{99} 
\bibitem{landau: 1975}
L. D. Landau and E. M. Lifshitz, The Classical Theory of Fields, Oxford: Pergamon Press (1975).

\bibitem{Arnowitt:1962hi} 
  R.~L.~Arnowitt, S.~Deser and C.~W.~Misner,
  Gen.\ Rel.\ Grav.\  {\bf 40}, 1997 (2008)
  [gr-qc/0405109].


\bibitem{Abbott:1981ff} 
  L.~F.~Abbott and S.~Deser,
  ``Stability of Gravity with a Cosmological Constant,''
  Nucl.\ Phys.\ B {\bf 195}, 76 (1982).


\bibitem{Deser:2002jk} 
  S.~Deser and B.~Tekin,
  ``Energy in generic higher curvature gravity theories,''
  Phys.\ Rev.\ D {\bf 67}, 084009 (2003)
  [hep-th/0212292].


\bibitem{Szabados:2004vb} 
  L.~B.~Szabados,
  ``quasi-local Energy-Momentum and Angular Momentum in GR: A Review Article,''
  Living Rev.\ Rel.\  {\bf 7}, 4 (2004).


\bibitem{Brown:1992br} 
  J.~D.~Brown and J.~W.~York, 
  ``Quasilocal energy and conserved charges derived from the gravitational action,''
  Phys.\ Rev.\ D {\bf 47}, 1407 (1993)
  [gr-qc/9209012].


\bibitem{Balasubramanian:1999re} 
  V.~Balasubramanian and P.~Kraus,
  ``A Stress tensor for Anti-de Sitter gravity,''
  Commun.\ Math.\ Phys.\  {\bf 208}, 413 (1999)
  [hep-th/9902121].


\bibitem{Komar:1958wp} 
  A.~Komar,
  ``Covariant conservation laws in general relativity,''
  Phys.\ Rev.\  {\bf 113}, 934 (1959).


\bibitem{Wald:1993nt} 
  R.~M.~Wald,
  ``Black hole entropy is the Noether charge,''
  Phys.\ Rev.\ D {\bf 48}, 3427 (1993)
  [gr-qc/9307038].

\bibitem{Jacobson:1993vj} 
  T.~Jacobson, G.~Kang and R.~C.~Myers,
  ``On black hole entropy,''
  Phys.\ Rev.\ D {\bf 49}, 6587 (1994)
  [gr-qc/9312023].

\bibitem{Iyer:1994ys} 
  V.~Iyer and R.~M.~Wald,
  ``Some properties of Noether charge and a proposal for dynamical black hole entropy,''
  Phys.\ Rev.\ D {\bf 50}, 846 (1994)
  [gr-qc/9403028].


\bibitem{Wald:1999wa} 
  R.~M.~Wald and A.~Zoupas,
  ``A General definition of 'conserved quantities' in general relativity and other theories of gravity,''
  Phys.\ Rev.\ D {\bf 61}, 084027 (2000)
  [gr-qc/9911095].


\bibitem{Bekenstein:1973ur} 
  J.~D.~Bekenstein,
  ``Black holes and entropy,''
  Phys.\ Rev.\ D {\bf 7}, 2333 (1973).


\bibitem{Barnich:2001jy} 
  G.~Barnich and F.~Brandt,
  ``Covariant theory of asymptotic symmetries, conservation laws and central charges,''
  Nucl.\ Phys.\ B {\bf 633}, 3 (2002)
  [hep-th/0111246].



\bibitem{Barnich:2003xg} 
  G.~Barnich,
  ``Boundary charges in gauge theories: Using Stokes theorem in the bulk,''
  Class.\ Quant.\ Grav.\  {\bf 20}, 3685 (2003)
  [hep-th/0301039].


\bibitem{Barnich:2004uw} 
  G.~Barnich and G.~Compere,
  ``Generalized Smarr relation for Kerr AdS black holes from improved surface integrals,''
  Phys.\ Rev.\ D {\bf 71}, 044016 (2005)
  [Erratum-ibid.\ D {\bf 71}, 029904 (2006)]
  [gr-qc/0412029].

\bibitem{Barnich:2007bf} 
  G.~Barnich and G.~Compere,
  ``Surface charge algebra in gauge theories and thermodynamic integrability,''
  J.\ Math.\ Phys.\  {\bf 49}, 042901 (2008)
  [arXiv:0708.2378 [gr-qc]].




\bibitem{Kim:2013zha} 
  W.~Kim, S.~Kulkarni and S.~-H.~Yi,
  ``quasi-local Conserved Charges in Covariant Theory of Gravity,''
  Phys.\ Rev.\ Lett.\  {\bf 111}, 081101 (2013)
  [arXiv:1306.2138 [hep-th]].


\bibitem{Majhi:2011ws} 
  B.~R.~Majhi and T.~Padmanabhan,
  ``Noether Current, Horizon Virasoro Algebra and Entropy,''
  Phys.\ Rev.\ D {\bf 85}, 084040 (2012)
  [arXiv:1111.1809 [gr-qc]].


\bibitem{Majhi:2012tf} 
  B.~R.~Majhi and T.~Padmanabhan,
  ``Noether current from the surface term of gravitational action, Virasoro algebra and horizon entropy,''
  Phys.\ Rev.\ D {\bf 86}, 101501 (2012)
  [arXiv:1204.1422 [gr-qc]].


\bibitem{Majhi:2012nq} 
  B.~R.~Majhi,
  ``Noether current of the surface term of Einstein-Hilbert action, Virasoro algebra and entropy,''
  Adv.\ High Energy Phys.\  {\bf 2013}, 386342 (2013)
  [arXiv:1210.6736 [gr-qc]].


\bibitem{Kim:2013qra} 
  W.~Kim, S.~Kulkarni and S.~-H.~Yi,
  ``Conserved quantities and Virasoro algebra in New massive gravity,''
  JHEP {\bf 1305}, 041 (2013)
  [arXiv:1303.3691 [hep-th]].


\bibitem{Deser:1981wh} 
  S.~Deser, R.~Jackiw and S.~Templeton,
  ``Topologically Massive Gauge Theories,''
  Annals Phys.\  {\bf 140}, 372 (1982)
  [Erratum-ibid.\  {\bf 185}, 406 (1988)]
  [Annals Phys.\  {\bf 185}, 406 (1988)]
  [Annals Phys.\  {\bf 281}, 409 (2000)].


\bibitem{Tachikawa:2006sz} 
  Y.~Tachikawa,
  ``Black hole entropy in the presence of Chern-Simons terms,''
  Class.\ Quant.\ Grav.\  {\bf 24}, 737 (2007)
  [hep-th/0611141].


\bibitem{Cho:2004ey} 
  J.~-H.~Cho,
  ``BTZ black-hole dressed in the gravitational Chern-Simons term,''
  J.\ Korean Phys.\ Soc.\  {\bf 44}, 1355 (2004).


\bibitem{Kraus:2005vz} 
  P.~Kraus and F.~Larsen,
  ``Microscopic black hole entropy in theories with higher derivatives,''
  JHEP {\bf 0509}, 034 (2005)
  [hep-th/0506176].


\bibitem{Kraus:2005zm} 
  P.~Kraus and F.~Larsen,
  ``Holographic gravitational anomalies,''
  JHEP {\bf 0601}, 022 (2006)
  [hep-th/0508218].


\bibitem{Solodukhin:2005ah} 
  S.~N.~Solodukhin,
  ``Holography with gravitational Chern-Simons,''
  Phys.\ Rev.\ D {\bf 74}, 024015 (2006)
  [hep-th/0509148].


\bibitem{Sahoo:2006vz} 
  B.~Sahoo and A.~Sen,
  ``BTZ black hole with Chern-Simons and higher derivative terms,''
  JHEP {\bf 0607}, 008 (2006)
  [hep-th/0601228].


\bibitem{Park:2006gt} 
  M.~-I.~Park,
  ``BTZ black hole with gravitational Chern-Simons: Thermodynamics and statistical entropy,''
  Phys.\ Rev.\ D {\bf 77}, 026011 (2008)
  [hep-th/0608165].
  
  
\bibitem{Bonora:2011gz} 
  L.~Bonora, M.~Cvitan, P.~Dominis Prester, S.~Pallua and I.~Smolic,
  ``Gravitational Chern-Simons Lagrangians and black hole entropy,''
  JHEP {\bf 1107}, 085 (2011)
  [arXiv:1104.2523 [hep-th]];
   
  L.~Bonora, M.~Cvitan, P.~Dominis Prester, S.~Pallua and I.~Smolic,
  ``Gravitational Chern-Simons Lagrangian terms and spherically symmetric spacetimes,''
  Class.\ Quant.\ Grav.\  {\bf 28}, 195009 (2011)
  [arXiv:1105.4792 [hep-th]];
  
  L.~Bonora, M.~Cvitan, P.~Dominis Prester, S.~Pallua and I.~Smolic,
  ``Gravitational Chern-Simons terms and black hole entropy. Global aspects,''
  JHEP {\bf 1210}, 077 (2012)
  [arXiv:1207.6969 [hep-th]];
  
  L.~Bonora, M.~Cvitan, P.~Dominis Prester, S.~Pallua and I.~Smolic,
  ``Stationary rotating black holes in theories with gravitational Chern-Simons Lagrangian term,''
  Phys.\ Rev.\ D {\bf 87}, 024047 (2013)
  [arXiv:1210.4035 [hep-th]].
  

\bibitem{Deser:2003vh} 
  S.~Deser and B.~Tekin,
  ``Energy in topologically massive gravity,''
  Class.\ Quant.\ Grav.\  {\bf 20}, L259 (2003)
  [gr-qc/0307073].

\bibitem{Bouchareb:2007yx} 
  A.~Bouchareb and G.~Clement,
  ``Black hole mass and angular momentum in topologically massive gravity,''
  Class.\ Quant.\ Grav.\  {\bf 24}, 5581 (2007)
  [arXiv:0706.0263 [gr-qc]].

\bibitem{Hotta:2008yq} 
  K.~Hotta, Y.~Hyakutake, T.~Kubota and H.~Tanida,
  ``Brown-Henneaux's Canonical Approach to Topologically Massive Gravity,''
  JHEP {\bf 0807}, 066 (2008)
  [arXiv:0805.2005 [hep-th]].
  
\bibitem{Chen:2013aza} 
  B.~Chen, J.~-j.~Zhang, J.~-d.~Zhang and D.~-l.~Zhong,
  ``Aspects of Warped AdS$_3$/CFT$_2$ Correspondence,''
  JHEP {\bf 1304}, 055 (2013)
  [arXiv:1302.6643 [hep-th]].


\bibitem{Anninos:2008fx} 
  D.~Anninos, W.~Li, M.~Padi, W.~Song and A.~Strominger,
  ``Warped AdS(3) Black Holes,''
  JHEP {\bf 0903}, 130 (2009)
  [arXiv:0807.3040 [hep-th]].


\bibitem{Guica:2008mu} 
  M.~Guica, T.~Hartman, W.~Song and A.~Strominger,
  ``The Kerr/CFT Correspondence,''
  Phys.\ Rev.\ D {\bf 80}, 124008 (2009)
  [arXiv:0809.4266 [hep-th]].

\bibitem{Strominger:2001pn} 
  A.~Strominger,
  ``The dS / CFT correspondence,''
  JHEP {\bf 0110}, 034 (2001)
  [hep-th/0106113].
  
\bibitem{Anninos:2009jt} 
  D.~Anninos,
  ``Sailing from Warped AdS(3) to Warped dS(3) in Topologically Massive Gravity,''
  JHEP {\bf 1002}, 046 (2010)
  [arXiv:0906.1819 [hep-th]].
  
\bibitem{deBuyl:2013ega} 
  S.~de Buyl, S.~Detournay, G.~Giribet and G.~S.~Ng,
  ``Baby de Sitter Black Holes and dS$_3$/CFT$_2$,''
  arXiv:1308.5569 [hep-th].



\bibitem{Banados:1992wn} 
  M.~Banados, C.~Teitelboim and J.~Zanelli,
  ``The Black hole in three-dimensional space-time,''
  Phys.\ Rev.\ Lett.\  {\bf 69}, 1849 (1992)
  [hep-th/9204099].



\bibitem{Moussa:2003fc} 
  K.~A.~Moussa, G.~Clement and C.~Leygnac,
  ``The Black holes of topologically massive gravity,''
  Class.\ Quant.\ Grav.\  {\bf 20}, L277 (2003)
  [gr-qc/0303042].


\bibitem{Deser:2002rt} 
  S.~Deser and B.~Tekin,
  ``Gravitational energy in quadratic curvature gravities,''
  Phys.\ Rev.\ Lett.\  {\bf 89}, 101101 (2002)
  [hep-th/0205318].


\bibitem{Senturk:2012yi} 
  C.~Senturk, T.~C.~Sisman and B.~Tekin,
  ``Energy and Angular Momentum in Generic F(Riemann) Theories,''
  Phys.\ Rev.\ D {\bf 86}, 124030 (2012)
  [arXiv:1209.2056 [hep-th]].


\bibitem{Amsel:2012se} 
  A.~J.~Amsel and D.~Gorbonos,
  ``A Wald-like Formula for Energy,''
  Phys.\ Rev.\ D {\bf 87}, 024032 (2013)
  [arXiv:1209.1603 [gr-qc]].





\bibitem{Nam:2010ub} 
  S.~Nam, J.~-D.~Park and S.~-H.~Yi,
  ``Mass and Angular momentum of Black Holes in New Massive Gravity,''
  Phys.\ Rev.\ D {\bf 82}, 124049 (2010)
  [arXiv:1009.1962 [hep-th]].


\bibitem{Lee:1990nz} 
  J.~Lee and R.~M.~Wald,
  ``Local symmetries and constraints,''
  J.\ Math.\ Phys.\  {\bf 31}, 725 (1990).
  
  
\bibitem{Borowiec:1998st} 
  A.~Borowiec, M.~Ferraris and M.~Francaviglia,
  ``Lagrangian symmetries of Chern-Simons theories,''
  J.\ Phys.\ A {\bf 31}, 8823 (1998)
  [hep-th/9801126];

  A.~Borowiec, M.~Ferraris and M.~Francaviglia,
  ``A Covariant formalism for Chern-Simons gravity,''
  J.\ Phys.\ A {\bf 36}, 2589 (2003)
  [hep-th/0301146].
  
  A.~Borowiec, L.~Fatibene, M.~Ferraris and M.~Francaviglia,
  ``Covariant Lagrangian formulation of Chern-Simons and BF theories,''
  Int.\ J.\ Geom.\ Meth.\ Mod.\ Phys.\  {\bf 3}, 755 (2006)
  [Int.\ J.\ Geom.\ Meth.\ Mod.\ Phys.\  {\bf 4}, 277 (2007)]
  [hep-th/0511060].

\bibitem{Nazaroglu:2011zi} 
  C.~Nazaroglu, Y.~Nutku and B.~Tekin,
  ``Covariant Symplectic Structure and Conserved Charges of Topologically Massive Gravity,''
  Phys.\ Rev.\ D {\bf 83}, 124039 (2011)
  [arXiv:1104.3404 [hep-th]].
  
  
\bibitem{Compere:2008cv} 
  G.~Compere and S.~Detournay,
  ``Semi-classical central charge in topologically massive gravity,''
  Class.\ Quant.\ Grav.\  {\bf 26}, 012001 (2009)
  [Erratum-ibid.\  {\bf 26}, 139801 (2009)]
  [arXiv:0808.1911 [hep-th]].

\bibitem{Ghodsi:2011ua} 
  A.~Ghodsi and D.~M.~Yekta,
  ``On Asymptotically AdS-Like Solutions of Three Dimensional Massive Gravity,''
  JHEP {\bf 1206}, 131 (2012)
  [arXiv:1112.5402 [hep-th]].
  
  
\bibitem{Carlip:1999cy} 
  S.~Carlip,
  ``Entropy from conformal field theory at Killing horizons,''
  Class.\ Quant.\ Grav.\  {\bf 16}, 3327 (1999)
  [gr-qc/9906126].


 
\end{thebibliography}
\end{document}